\documentclass[prb,aps,amssymb,floatfix,twocolumn,showpacs,preprintnumbers]{revtex4}
\usepackage{amssymb}
\usepackage{amsfonts}
\usepackage{amsmath}
\usepackage{graphicx}

\def\be{\begin{equation}}
\def\ee{\end{equation}}
\def\bea{\begin{eqnarray}}
\def\eea{\end{eqnarray}}

\def\wc{{\omega_c}}

\begin{document}

\title{Negative conductivity and anomalous screening in two-dimensional electron systems
subjected to microwave radiation}
\author{S.~I.~Dorozhkin,$^{1}$  I.~A.~Dmitriev,$^{2,3}$ and A.~D.~Mirlin$^{2,4,5}$}
\affiliation{$^1$Institute of Solid State Physics, Chernogolovka,
Moscow district, 142432, Russia\\
$^{2}$Institut f\"ur
Nanotechnologie, Karlsruhe Institute of Technology, 76021 Karlsruhe, Germany\\
$^{3}$Ioffe Physical Technical Institute, 194021 St.~Petersburg, Russia\\
$^{4}$Institut f\"ur Theorie der kondensierten Materie and DFG Center for Functional Nanostructures,
Karlsruhe Institute of Technology, 76128 Karlsruhe, Germany\\
$^{5}$Petersburg Nuclear Physics Institute, 188300
  St. Petersburg, Russia
}

\begin{abstract}
A 2D electron system in a quantized magnetic field can be driven
by microwave radiation into a non-equilibrium state with strong
magnetooscillations of the dissipative conductivity. We
demonstrate that in such system a negative conductivity can
coexist with a positive diffusion coefficient. 
In a finite system, solution of coupled electrostatic 
and linear transport problems shows that the diffusion can
stabilize a state with negative conductivity. Specifically, this
happens when the system size is smaller than the absolute value of
the non-equilibrium screening length that diverges at the point
where the conductivity changes sign. We predict that a negative
resistance can be measured in such a state. Further, for a
non-zero difference between the work functions of two contacts, 
we explore the distribution of the
electrostatic potential and of the electron density in the sample.
We show that in the diffusion-stabilized regime of negative
conductivity the system splits into two regions with opposite
directions of electric field. This effect is a precursor of the
domain structure that has been predicted to emerge spontaneously
in the microwave-induced zero-resistance states.

\end{abstract}

\date{\today}
\maketitle

\section{Introduction}
High mobility two-dimensional electron systems (2DES) subjected to
the microwave radiation reveal giant magnetooscillations in the diagonal
resistance $\rho_{\rm xx}$~\cite{Zudov1,Ye} with periodicity controlled 
by the positions
of the cyclotron resonance harmonics. In high-quality samples,
these oscillations give rise to zero resistance
states~\cite{Mani,Zudov2,Zudov3} (ZRS) at the oscillation minima, 
$\rho_{\rm xx}\to 0$,
when the radiation power is increased and the temperature is
lowered. In the Corbino-disk geometry ZRS manifest themselves as
zero-conductance states. The widely accepted explanation of ZRS
combines two distinct phenomena. The first one is development of
the giant magnetoresistance oscillations due to indirect optical
transitions~\cite{Ryzhii,Durst,Vavilov} or nonequilibrium
occupation of electronic
states~\cite{Dorozhkin1,dmitriev03,Dmitriev1}. Both mechanisms
predict, under appropriate conditions, a negative resistivity
value in the oscillation minima at a small dc current. At some
finite current density $j_0$, the resistivity crosses zero and
becomes positive~\cite{Andreev,Vavilov,Dmitriev1}. The second
effect invoked for explanation of ZRS is spontaneous formation of
current domains with current density $j_0$~\cite{Andreev} which
occurs due to instability of systems with negative absolute and/or
differential resistivity~\cite{Zakharov,domains-books,Andreev}. In
the simplest geometry the system breaks into two domains carrying
equal Hall currents of opposite direction. The Hall electric
fields related to these currents are also equal in magnitude and
have opposite directions. The resulting dissipative component of
the resistivity or conductivity tensor is zero \cite{Andreev}. Two
works \cite{willett04,dorozhkin11} provided an experimental
evidence supporting the spontaneous domain formation in ZRS.

In this paper we show that a positive diffusion coefficient can stabilize
a homogeneous state
of a {\it finite} 2DES even though the system has a negative resistivity.
We point out that combination of a negative resistivity and a positive
diffusion coefficient naturally arises in 2DES under microwave radiation.
Solving the transport equation jointly with the electrostatic
problem for a finite 2DES, we determine a stability condition which
includes conductivity, diffusion coefficient, and the system size.

Further, we allow for a non-zero difference between the work
functions of two contacts to 2DES (leads), that gives rise to the
photogalvanic effects in an irradiated sample \cite{Dmitriev2}. We
explore the distribution of the electrostatic potential and the
electron density in such a sample for different values of the
conductivity. Most remarkably, we find that in the
diffusion-stabilized regime of negative conductivity the system
splits into two regions with opposite directions of electric
field. We argue that this effect is a precursor of the  domain structure that is 
expected to emerge spontaneously in microwave-induced ZRS.

\section{Model and basic results}

\subsection{Microscopic conductivity and diffusion coefficient}
The local dissipative current density in a weakly non-uniform 2DES with
surface electron density $n_s$ has a form
\begin{equation}
j=-2\sigma\nabla_s \phi|_{z=0} -e D\nabla_s {n_s}\,,
\label{j}
\end{equation}
where $\phi|_{z=0}$ is the electrical potential in the 2DES plane $z=0$,
$\nabla_s$ is the surface gradient in this plane, $\sigma$ is the
dissipative component of the magnetoconductivity tensor (per
spin), and $D$ is the diffusion coefficient.

In equilibrium, the conductivity $\sigma_0$ and the diffusion
coefficient $D_0$ obey the Einstein relation $2\sigma_0=e^2 \chi
D_0$, where $\chi=\partial n_s/\partial\mu$ is the equilibrium
static compressibility (here $\mu$ is the 2DES chemical
potential). The Einstein relation guarantees that $j=-2\sigma
\nabla_s \eta/e$ is zero in the equilibrium state with constant
electrochemical potential $\eta=e\phi|_{z=0}+\mu$; further, the
current response does not depend on the type of perturbation
(electric field or density gradient).

In the presence of microwaves the Einstein relation is violated,\cite{Dmitriev2,kashuba06}
\begin{equation}
2\sigma\neq e^2 \chi D\,.
\label{Einstein}
\end{equation}
This violation leads to the photogalvanic effects observed
experimentally in Refs.~\onlinecite{willett04,Dorozhkin3} and is
at the heart of the phenomena discussed in the present work.

\begin{figure}[tb]
\includegraphics[width=\columnwidth]{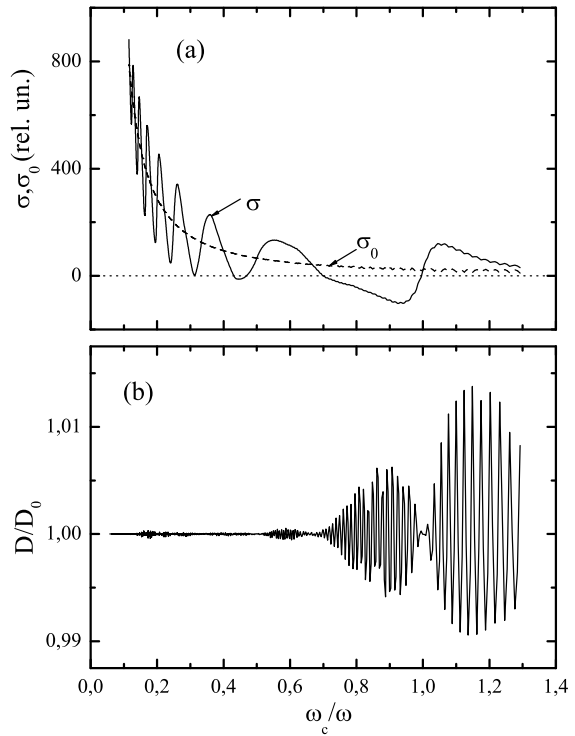}
\caption{Magnetic-field dependence of (a) the dissipative
conductivity $\sigma_0$, $\sigma$  and (b) the normalized electron
diffusion coefficient $D/D_0$, shown as a function of the ratio
$\omega_{\rm c}/\omega$, where $\omega_{\rm c}$ is the electron
cyclotron frequency and $\omega$ is the circular radiation
frequency. The quantities $\sigma_0$ and $D_0$ correspond to the
dark conditions, while $\sigma$ and $D$ were calculated for a
nonequilibrium state under microwave radiation. Parameters of
calculations: electron surface density $n_{\rm s}$ corresponds to
the filling factor of the Landau levels $N\approx 100$ at
$\omega_{\rm c}/\omega=1$, $kT/\hbar\omega=0.2$, and
$\omega\tau_{\rm q}=10$. Here $T$ is the dark temperature of 2DES
and $\tau_{\rm q}$ is the quantum scattering time determining the
Landau level broadening.} \label{fig1}
\end{figure}

We start with demonstration that the theory~\cite{Dmitriev2} allows
for coexistence of a negative dissipative conductivity
and a positive diffusion coefficient. This
result is the most prominent for the case when effect of the
microwave radiation on the electron kinetics is governed by
modification of the electron distribution function
(inelastic mechanism).\cite{comment}
For this mechanism, equations for the conductivity $\sigma$ and
the diffusion coefficient $D$ read
\begin{equation}
\sigma=-\sigma_D\int\frac{\nu^2(\varepsilon)}{\nu_0^2}
\frac{\partial f(\varepsilon)}{\partial \varepsilon}\,d\varepsilon,
\label{sigma}
\end{equation}
\begin{equation}
D=\frac{2 \sigma_D}{e^2}\int\frac{\nu^2(\varepsilon)}{\nu_0^2}
\frac{\partial f(\varepsilon)}{\partial n_{\rm s}}\,d\varepsilon.
\label{D}
\end{equation}
Here $\sigma_D=n_{\rm s}e^2/2m^*\omega_{\rm c}^2\tau$ is the Drude
conductivity in a classically strong magnetic field, $\omega_{\rm
c}\tau \gg 1$; $m^*$ is the effective mass, $\omega_{\rm c}$ the
cyclotron frequency, and $\tau$ the transport relaxation time. We
remind the reader that at $\omega_{\rm c}\tau \gg 1$ the
dissipative resistivity is proportional to $\sigma$ (since the
dominant component of the conductivity tensor is the Hall
conductivity that is only weakly affected by microwaves). Further,
$\nu(\varepsilon)$ is the density of states in disorder-broadened
Landau levels and $\nu_0=m^*/2\pi\hbar^2$ is the density of states
per spin at $B=0$. The non-equilibrium distribution function
$f(\epsilon)$ is determined by the kinetic equation.\cite{Dmitriev1}
Figure~\ref{fig1} illustrates the magnetic-field
dependence of the dissipative conductivity $\sigma$ and of the
diffusion coefficient $D$ for typical parameters. (Details of
numerical procedure used for calculation of density of states
$\nu(\varepsilon)$ and nonequilibrium distribution function
$f(\varepsilon)$ are given in Ref.~\onlinecite{Dorozhkin2}.) Under
microwave radiation with the circular frequency $\omega$, $\sigma$
shows strong magnetooscillations (with periodicity controlled by
the ratio $\omega/\omega_{\rm c}$) and becomes negative around
minima. At the same time, $D$ remains almost unaffected by
microwaves (apart from Shubnikov--de Haas oscillations that are
strongly suppressed due to temperature smearing; for parameters in
Fig.~\ref{fig1} their amplitude remains within 1\%). 
%In what follows we will neglect small variations of $D$, assuming $D = D_0> 0$.

%%%%%%%%%%%%%%%%%%%%%%%%%%%%%%%%%%%%%%%%%%%%%%%%%%%%%%%%%%%%%%%%%%%%%%%%%%%%%%%%%%%%%%
%%%%%%%%%%%%%%%%%%%%%%%%%%%%%%%%%%%%%%%%%%%%%%%%%%%%%%%%%%%%%%%%%%%%%%%%%%%%%%%%%%%%%%

\subsection{Electrostatics and transport in a 2D stripe}
%%%%%%%%%%%%%%%%%%%%%%%%%%%%%%%%%%%%%%%%%%%%%%%%%%%%%%%%%%%%%%%%%%%%%%%%%%%%%%%%%%%
\begin{figure}[tb]
\includegraphics[width=0.6\columnwidth]{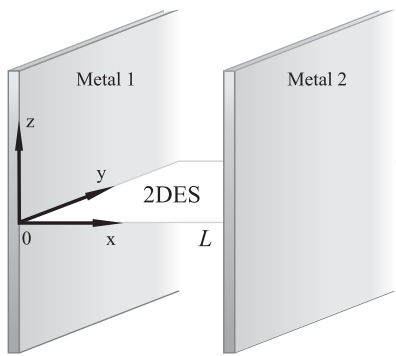}
\caption{Schematic view of the 2DES and metal contacts for
considered electrostatic problem.} \label{fig2}
\end{figure}
%%%%%%%%%%%%%%%%%%%%%%%%%%%%%%%%%%%%%%%%%%%%%%%%%%%%%%%%%%%%%%%%%%%%%%%%%%%%%%%%%%%%%%

We proceed now by finding a self-consistent solution of
electrostatic and transport problems in a sample of finite length
$L$ subjected to microwave radiation and dc electric field. The
considered geometry is shown in Fig.~\ref{fig2}. Specifically, we
assume that 2DES forms a stripe in the $(x,y)$ plane between two
infinite metal plates at $x=0$ and $x=L$ which model Ohmic
contacts. The magnetic field $B$ is parallel to the $z$ axis. In
order to find a stationary spatial distribution of the 2D electron
density $n_s(x)$ and of the potential $\phi(x,z)$ we use a system
of coupled equations that includes the transport equation
(\ref{j}), the continuity equation 
\be\label{con}
e\dot{n}_s+\nabla_s\cdot {\bf j}=0, 
\ee 
and the Poisson equation
\be\label{poisson} 
-\epsilon\Delta\phi(x,z)=4\pi e
(n_s-n_+)\delta(z). 
\ee 
Here $\epsilon$ is the dielectric constant
of the medium surrounding 2DES, and a constant $n_+$ represents
the frozen positive background charge of ionized donors. 
In the bulk of an infinite sample, the electroneutrality requires
$n_s=n_+$. In what follows, we assume
that relative density variations are weak, $n_s-n_+\ll n_+$, which
allows us to use spatially-independent $\sigma$ and $D$ in
Eq.~\eqref{j}. For simplicity, the system is assumed to be
infinite in $y$ direction. This setup can be termed
``quasi-Corbino'', with $x$ axis corresponding to the radial
direction and $y$ to the azimuthal direction of the Corbino disk.
By symmetry, $n_s(x)$ and $\phi(x,z)$ do not depend on $y$, and
the dissipative current \eqref{j} flows parallel to
$x$--direction. The local density and electric field define the
Hall component $j_y(x)=e n(x) v_D(x)$, a dissipationless flow in
$y$--direction with local drift velocity $v_D=-(c/B)\partial
\phi(x,0)/\partial x$.

The above equations are supplemented by boundary conditions for the electrostatic potential,
\bea
\label{BCphi1}
&& \phi(0,z)=\phi_0\,, \qquad \phi(L,z)=0\,, \\
&& (\partial\phi/\partial z)_{z\to\pm\infty}\to 0\,, \label{BCphi2}
\eea
and for the density,
\be\label{BCn}
n_s(L)=n_+\,,\quad n_s(0)=n_+ -\chi e{\cal U}_c\,.
\ee
In Eq.~(\ref{BCn}), $e{\cal U}_c$
is the difference of work functions between
the left lead and 2DES. Due to large density of states in the
metallic lead, the interface charge density in 2DES
$e(n_s(0)-n_+)$ is fixed by $e{\cal U}_c$ and is not affected by
either the radiation or the current flow (see
Ref.~\onlinecite{Dmitriev2} for details). For simplicity, we
assume that the right lead has the same work function as 2DES, so
that $n_s(L) = n_+$. As we show below, a finite $e{\cal U}_c\neq 0$
makes the electrostatic problem nontrivial by introducing the spatial 
variation of the density $n_s$.
 Integrating Eq.~(\ref{j}) over $x$ from $0$ to $L$, we get
a relation between the current density $j$, the electrostatic potential
difference $\phi_0$, and the chemical potential difference $e{\cal
U}_c$, \cite{Dmitriev2} 
\be \label{j-phi-uc} 
jL = 2\sigma\phi_0 -e^2\chi D {\cal U}_c.
\ee

In the absence of 2DES ($n_s-n_+=0$) the Poisson equation
\eqref{poisson} has a trivial solution $\phi(x,z) = \phi_0(1-x/L)$.
Splitting this contribution and
expanding the remaining part in eigenfunctions of the Laplace operator
with proper boundary conditions [Eqs.~(\ref{BCphi1}) and
(\ref{BCphi2}) with $\phi_0=0$], we get
\begin{equation}
\phi(x,z)=\phi_0\left(1-\frac{x}{L}\right)+\sum_{k=1}^\infty A_{\rm
k}e^{-k\pi |z|/L}\sin \frac{k\pi x}{L}\,.
\label{phi_gen}
\end{equation}
With the use of Eq.~(\ref{poisson}), the charge density can be
expressed in terms of coefficients $A_{\rm k}$ as follows:
\be\label{nk} 
e(n_s(x)-n_+)=\sum_{k=1}^\infty \frac{\epsilon
k}{2L} A_k\sin \frac{\pi k x}{L}. 
\ee 
We integrate now
Eq.~(\ref{j}) from $x$ to $L$, 
\be \label{j_int} 
j(L-x) = 2\sigma
\phi(x,0) +eD [n_s(x) - n_+]\,, 
\ee 
and substitute here Eqs.~\eqref{phi_gen} and \eqref{nk}. Using 
\be
\label{fourier-linear} 
1-\frac{x}{L} = \frac{2}{\pi}
\sum_{k=1}^\infty k^{-1} \sin {\frac{\pi k x}{L}}\,, 
\ee 
we finally obtain the coefficients $A_{\rm k}$ governing the
stationary solution of the problem:
\be
\label{Ak}
A_k=-\frac{4L}{\pi\epsilon k}\;\frac{e^2 \chi {\cal
U}_c}{k+L/\pi\lambda}\,.
\ee
Here we employed the relation
(\ref{j-phi-uc}) and introduced the non-equilibrium screening
length 
\be\label{lambda} 
\lambda=\frac{\epsilon D}{4\pi\sigma} 
\ee
(discussed in more detail below).

\subsection{Nonequilibrium screening length}

The nonequilibrium screening length  (\ref{lambda}) which enters
the solution \eqref{Ak}, plays a central role in nonequilibrium
transport in inhomogeneous 2DES. In equilibrium (in the absence of
microwaves), the Einstein relation holds and $\lambda$ reduces to
the conventional 2D Thomas-Fermi screening length,
\be\label{lambda0} 
\lambda\to\lambda_0=\frac{\epsilon }{2\pi
e^2\chi},\qquad 2\sigma_0=e^2\chi D_0. 
\ee

Using Eqs.~\eqref{j}, \eqref{con}, and \eqref{poisson}, it is easy
to show that indeed $\lambda$ replaces the equilibrium $\lambda_0$
in all electrostatic problems. Under microwave radiation, the local
conductivity $\sigma$ in Eq.~\eqref{lambda} can cross zero and become
negative, as illustrated in
Fig.~\ref{fig1}. The screening length $\lambda$ diverges at $\sigma = 0$.
As we show below, in the regime
$\sigma<0$ the stationary solutions of the linear problem become electrically unstable on
the spatial scale $|\lambda|$ determined by the nonequlibrium
screening length. 

\subsection{Stability conditions: finite-size effects}
\label{stability}
We check now the stability of the obtained solution with respect to
slow spatio-temporal fluctuations. To this end, we add a fluctuating
part $\tilde{n}$, $\tilde{\phi}$ and $\tilde{j}$ to $n_s$, $\phi$,
and $j$. To satisfy the boundary conditions
$\{\tilde{n},\tilde{\phi}\}|_{x=0,L}=0$,  we take 
\be\label{deltan}
\tilde{n}(x,y,t)=\delta n_q(t) e^{i q_y y} \sin q_x x 
\ee 
with
$q_x=\pi k/L$, $k=1,2,..$, and continuous $q_y$. Using
Eq.~\eqref{poisson}, the corresponding 
\be\label{deltaphi}
\tilde{\phi}(x,y,z,t)=\frac{2\pi e}{\epsilon q} \delta n_q(t) e^{i
q_y y-q|z|}\sin q_x x\,, 
\ee 
where $q=(q_x^2+q_y^2)^{1/2}$. The
fluctuating part of the current is therefore 
\be\label{tildej}
\tilde{j}=-2\sigma\nabla_s\tilde{\phi}|_{z=0}-eD\nabla_s\tilde{n}-2\sigma_H\hat{\epsilon}\nabla_s\tilde{\phi}|_{z=0}\,,
\ee 
where we added to Eq.~(\ref{j}) the Hall term with $\sigma_H=en_s c/B$ and
$\epsilon_{xy}=-\epsilon_{yx}=1$. The continuity equation
\eqref{con} gives 
\be\label{con1} 
\frac{\partial\tilde{n}}{\partial
t}=\frac{2\sigma}{e}\Delta_s\tilde{\phi}|_{z=0}+D\Delta_s\tilde{n}=-\left(\frac{4\pi
\sigma}{\epsilon q}+D\right)q^2 \tilde{n}. 
\ee 
(The Hall term drops
out since $\nabla_s\hat{\epsilon}\nabla_s\tilde{\phi}=0$.)
Therefore, fluctuations do not grow in time if the stability
condition 
\be\label{stab} 
\lambda^{-1}+q>0 
\ee 
is satisfied. The
condition is most restrictive for soft modes with small $q$. In
an infinite system, where perturbations at arbitrarily long spatial
scale $q^{-1}$ are possible, the condition \eqref{stab} reduces to
the usual one: $\lambda>0$ or, equivalently,
$\sigma>0$.\cite{Andreev} In any finite system, $q^{-1}$ is limited
by the system size, and instability threshold shifts to negative
$\sigma$. In particular, in 2D stripe of width $L$, the minimal
wavenumber $q=\pi/L$ corresponds to the lowest harmonics $k=1$ in
Eqs.~\eqref{phi_gen}-\eqref{Ak}, and the stability condition
\eqref{stab} reads 
\be\label{stab1} 
L/\pi\lambda>-1,
\ee 
or, equivalently, 
\be\label{stab1a} 
\sigma>-\epsilon D/ 4 L. 
\ee

\section{Analysis of results}
In this section we discuss the obtained result \eqref{Ak}
for the field \eqref{phi_gen} and density \eqref{nk}
distribution in different physical situations.

\subsection{Homogeneous stable state with negative conductivity}
\label{homo} 
In the plain-capacitor contact configuration,
Fig.~\ref{fig2}, and for vanishing difference of the contact work functions, ${\cal
U}_c=0$, all harmonics $A_k=0$, see Eq.~\eqref{Ak}. The electron
density \eqref{nk} in initially homogeneous 2DES remains constant,
$n_{\rm s}=n_+$, independent on external bias $e
V\equiv\eta(0)-\eta(L)=e\phi(0)-e\phi(L)$, and
$\phi(x,z)=V(1-x/L)$, see Eq.~\eqref{phi_gen}. The dissipative
current \eqref{j} reduces to 
\be\label{miro} 
j=2 \sigma V/L,
\ee 
where $\sigma$ manifests the microwave-induced oscillations
illustrated in Fig.~\ref{fig1}a.

In the finite system,
the result (\ref{miro}) holds even for negative conductivity values, 
as long as the stability condition (\ref{stab1a}) is satisfied.
Our analysis in Sec.~\ref{stability} shows that diffusion with $D>0$ can stabilize
otherwise unstable\cite{Andreev} homogeneous state with $\sigma<0$
in a finite 2DES. According to Eq.~\eqref{stab1}, the lowest
observable value of negative conductivity in a stable homogeneous
state is 
\be\label{sigma_c} 
\sigma_c=-\epsilon D/4L\simeq
-\pi\lambda_0\sigma_0/L, 
\ee 
where in the last equality we used
$D\simeq D_0$ and Eq.~\eqref{lambda0}; the subscript $0$ refers to
the dark equilibrium state. The critical value $\sigma_c$
corresponds to the critical value $\lambda_c^{-1}=-\pi/L$ of the
inverse nonequilibrium screening length \eqref{lambda} fixed by
the lowest possible wavevector in the system.

At $\sigma<\sigma_c<0$, the system becomes electrically unstable and
breaks into domains. Possible domain configurations and their
dynamics were addressed in
Refs.~\onlinecite{{volkov:2004,Halperin05,Balents05,Halperin09}}.
The appearance of electric domains implies an accumulation of
charge at the boundary between domains of opposite polarity
(domain walls). The nonequilibrium screening length sets the
spatial scale of modulation of electronic density and
electrostatic potential in the domain phase. Therefore, a proper
account for the violated Einstein relation is crucial for
understanding the critical properties of the nonequilibrium phase
transition to ZRS, the microscopic structure of the domain walls,
their dynamics, and sensitivity to boundary conditions and details
of the disorder potential.

It should be mentioned that a negative residual conductivity has also
been found in Ref.~\onlinecite{volkov:2004}. Specifically, that work considered
the domain phase ($\sigma<\sigma_c$ in our terminology) and obtained
an exponentially small negative residual conductivity.
We believe that the exponential dependence found in
Ref.~\onlinecite{volkov:2004} is most likely an artefact of local
electrostatic approximation used in
Refs.~\onlinecite{volkov:2004,Halperin05,Balents05,Halperin09}.
Such an approximation
is appropriate for description of, e.g., the Gunn effect in bulk 3D
semiconductors, but is not directly applicable to 2DES
with quantized motion in $z$-direction. In contrast to 3D geometry,
in 2D the relation between $\phi$ and $n_s$ is non-local,
see Eq.~\eqref{poisson}. For proper description of the
domain phase, one should go beyond the linear response
in Eq.~\eqref{j}. This regime requires development of
adequate approaches for solution of arising nonlinear nonlocal
equations and will be addressed elsewhere.

\subsection{Photogalvanic effects}
We return now to an inhomogeneous system with a difference of work
functions in 2DES and the left lead $e{\cal U}_c<0$. For
simplicity, we also assume that temperature is sufficiently high
to suppress the magnetooscillations of the dark compressibility $\chi$
and Shubnikov-de Haas oscillations, $2\pi^2 kT/\hbar \wc\gg 1$. 
In this case, $D\simeq D_0$ (see Fig.~\ref{fig1}b) and
$\chi=\partial n_s/\partial \mu= n_s/\mu=2\nu_0$, so that the
electrochemical potential $\eta=e\phi|_{z=0}+n_s/2\nu_0$.

\subsubsection{Current-voltage characteristics}
The knowledge of spatial distributions \eqref{phi_gen} and
\eqref{nk} is not required for calculation of the current-voltage
characteristics (CVC). Indeed, the CVC is given by the relation
\eqref{j-phi-uc} which is obtained directly from Eq.~\eqref{j}
using boundary conditions \eqref{BCn}. The electrostatic potential
$\phi_0\equiv\phi(0,0)$ can be expressed 
from Eqs.~\eqref{BCphi1} and \eqref{BCn} in terms of 
electrochemical potential drop $eV\equiv\eta(0)-\eta(L)$ 
(measurable voltage across the sample) as
$\phi_0=V+{\cal U}_c$. Therefore, the CVC \eqref{j-phi-uc} reads
\be\label{CVC} 
j=2 \sigma \frac{V}{L}+2 \sigma\frac{{\cal
U}_c}{L}\left(1-\frac{\lambda}{\lambda_0}\right). 
\ee 
The last
term is responsible for photogalvanic effects. In the presence of
contact asymmetry ${\cal U}_c\neq 0$ or other source of the
built-in electric field ${\cal U}_c/L$ and provided the Einstein
relation is violated, $\lambda\neq\lambda_0$, the system displays
photocurrent $j\neq 0$ at zero  bias voltage $V=0$, and
photovoltage $V\neq 0$ at $j=0$, see Ref.~\onlinecite{Dmitriev2}
for details. Our present results show that this characteristic
stays valid for $\sigma<0$ if the condition \eqref{stab1} is
fulfilled. For the case of interest, $D>0$,
Eq.(\ref{CVC}) predicts that both the differential resistance
$dV/dj$ and the photo-voltage $V_{\rm photo}={\cal
U}_c(\lambda/\lambda_0-1)$ change sign when conductivity goes
through the zero value. This prediction can be verified
experimentally, e.g. on narrow Corbino disk samples which should
behave similarly to the infinite stripe considered here. 
Importantly, the size of the sample $L/\lambda_0$ should not be
very large in order to enable controllable measurements in the
stable region \eqref{stab1} with $\sigma<0$.

\subsubsection{Field and density distributions}
%Despite in the linear response regime and for simple 1D geometry we used
%the boundary values of $\phi$ and $n_s$ fully determine the CVC \eqref{CVC},
%in general
The charge and field distribution in the interior
of the sample provides an additional
insight into the problem and uncovers an interesting behavior in the
vicinity of the instability threshold. Using Eqs.~\eqref{phi_gen}-\eqref{Ak}
and \eqref{CVC} we obtain the density profile
\bea\label{n}
&&\frac{n_s(x)-n_+}{n_+-n_s(0)}\equiv\frac{\mu(x)-\mu(L)}{e\,{\cal U}_c}
=-{\cal N}\left(\frac{\pi x}{L},\frac{L}{\pi\lambda}\right)\!,\;\;
\eea
where
\bea
\label{N}
&&{\cal N}(X,\varkappa)=\frac{2}{\pi}\sum_{k=1}^\infty \frac{\sin kX}{k+\varkappa}
=\frac{2}{\pi}{\rm Im}\,\Phi(e^{iX},1,\varkappa),\;\;
\eea
and $\Phi(r,s,v)=\sum_{l=0}^\infty(v+l)^{-s}r^l$ is the Lerch transcendent.
% (GR 9.550).
Further, the 3D electrostatic potential reads 
\bea\nonumber 
\phi(x,z)&=&(V+\,{\cal U}_c)\left(1-\frac{x}{L}\right)\\
&-&
\frac{2L\,{\cal U}_c}{\pi^2\lambda_0}
\sum_{k=1}^\infty \frac{\sin (\pi k x/L)}{k+L/\pi\lambda}\,
\frac{e^{-\pi k |z|/L}}{k}\,. 
\label{phi}
\eea
At $z=0$, %, using $\alpha/k(k+\alpha)=1/k-1/(k+\alpha)$,
%the electrostatic potential
it can be represented as
\bea
\nonumber
\phi(x,0)&=&\left[V+\,{\cal U}_c\left(1-\frac{\lambda}{\lambda_0}\right)\right]\left(1-\frac{x}{L}\right)
\\&+&{\cal U}_c\frac{\lambda}{\lambda_0}{\cal N}\left(\frac{\pi x}{L},\frac{L}{\pi\lambda}\right)\,.
\label{phiPhi}\eea
Finally, %for
the electrochemical potential
%$\eta(x)=e\phi(x,0)+\mu(x)$ we get
 \bea
\nonumber
&&\eta(x)-\eta(L)=eV\left(1-\frac{x}{L}\right)
\\&&+e\,{\cal U}_c\left(1-\frac{\lambda}{\lambda_0}\right)\left[1-\frac{x}{L}
-{\cal N}\left(\frac{\pi x}{L},\frac{L}{\pi\lambda}\right)\right]\,.
\label{eta}\eea

\subsubsection{Enhancement of built-in field}
\label{enhancement}

%%%%%%%%%%%%%%%%%%%%%%%%%%%%%%%%%%%%%%%%%%%%%%%%%%%%%%%%%%%%%%%%%
\begin{figure}[tb]
\includegraphics[width=\columnwidth]{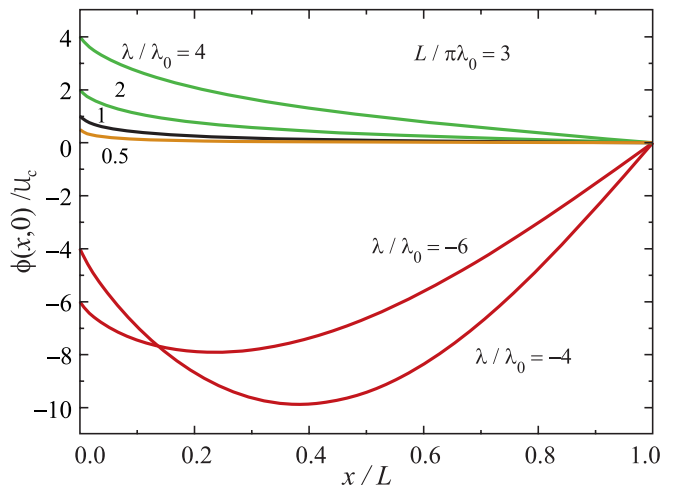}
\caption{The electrostatic potential $\phi(x,0)/{\cal U}_c$ vs.
 $x/L$ for $j=0$, $L=3\pi\lambda_0$, and several different $\lambda/\lambda_0$,
see Eq.~\eqref{phiPhi}.
Four upper curves correspond to $\sigma>0$ ($\lambda>0$). Two lower
curves are calculated for $\sigma<0$ satisfying the
stability condition $L/\pi\lambda> -1$.}
\label{fig3}
\end{figure}
%%%%%%%%%%%%%%%%%%%%%%%%%%%%%%%%%%%%%%%%%%%%%%%%%%%%%%%%%%%%%%%%%%%%%

Assuming an open circuit, $j=0$, in Fig.~\ref{fig3} we show 
how the built-in electrostatic potential
$\phi(x,0)$ is modified through the microwave-induced changes of
the screening length $\lambda$. Equation \eqref{CVC} for the case
$j=0$ yields 
\be\label{j=0} 
\phi(0,0)/{\cal U}_c= V/{\cal
U}_c+1=\lambda/\lambda_0 \simeq \sigma_0/\sigma\,. 
\ee 
In the last equality, we used $D\simeq D_0$. The electrostatic potential drop along the
sample is enhanced at $0<\sigma<\sigma_0$ (i.e. $\lambda/\lambda_0>1$), which is illustrated by two upper (green) curves
in Fig.~\ref{fig3} (the colors in Fig.~\ref{fig3} and \ref{fig4} refer to the online version).
The black curve (third from above) illustrates the equilibrium distribution. 
In the case $\sigma>\sigma_0$ (orange curve $\lambda/\lambda_0=0.5$ in Fig.~\ref{fig3}), 
the potential drop diminishes with respect to the equilibrium case.
This behavior at $\sigma>0$ illustrates the enhancement 
($\sigma<\sigma_0$) and suppression ($\sigma>\sigma_0$) of the built-in
electric field proposed for interpretation of photogalvanic effects 
in Ref.~\onlinecite{Dorozhkin3} on the experimental basis.

At $\sigma\to 0$, the potential $\phi_0\equiv\phi(0,0)$ diverges in an open circuit,
(namely, $\phi_0\to +\infty$ at $\sigma\to +0$ and $\phi_0\to -\infty$ at $\sigma\to
-0$). This divergence is cut by nonlinear corrections to $\sigma$ and $D$
which are not taken into account here.\cite{footnote1} 
At $\sigma<0$ and $j=0$, 
illustrated by two lowest (red) curves in Fig.~\ref{fig3}, the
potential $\phi(x,0)$ develops a single minimum which approaches
the position $x=L/2$ when $\sigma$ approaches the critical value
$\sigma_c = -\pi\lambda_0\sigma_0/L$, Eq.~\eqref{sigma_c}. In
Fig.~\ref{fig3} we took $L=3\pi\lambda_0$, so the corresponding critical
value of $\lambda$ is $\lambda_c=-L/\pi=-3\lambda_0$.

\subsubsection{Analysis of the field and density distributions in different regimes}
\label{regimes}

\begin{figure}[tb]
\includegraphics[width=\columnwidth]{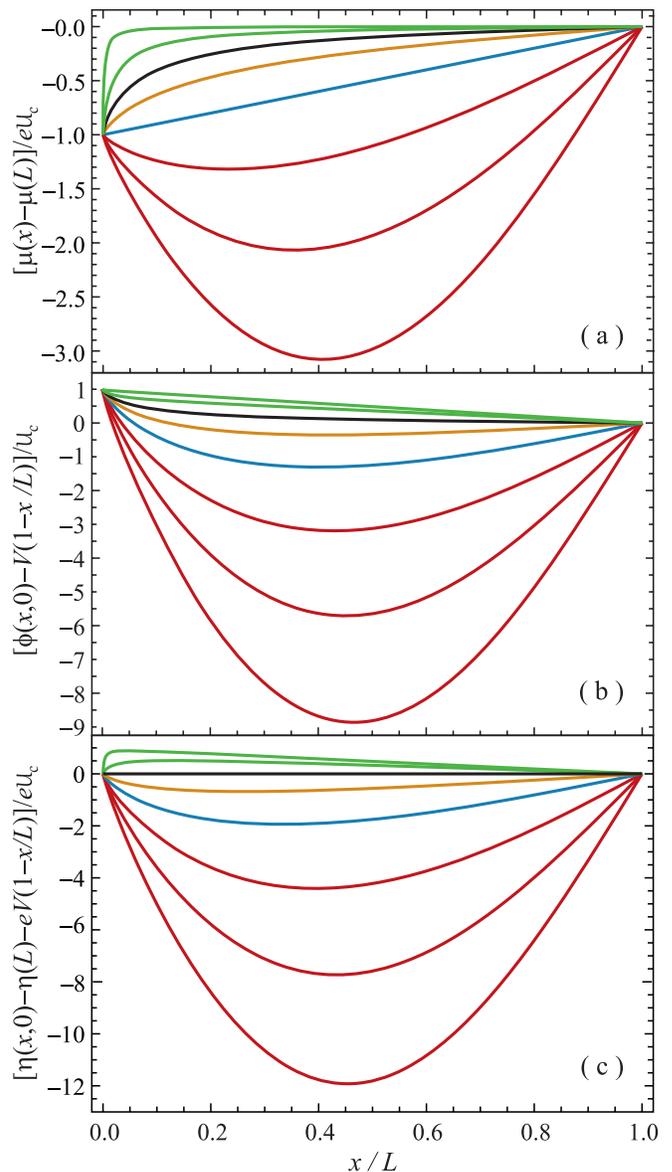}\\
\caption{Spatial distribution of (a) charge density, $[\mu(x)-\mu(L)]/e\,{\cal U}_c$ vs.
$x/L$, Eq.~(\ref{n});
(b) the electrostatic potential, $[\phi(x,0)-V(1-x/L)]/\,{\cal U}_c$ vs. $x/L$, Eq.~(\ref{phiPhi});
and (c) the electrochemical potential, $[\eta(x)-\eta(L)-eV(1-x/L)]/e\,{\cal U}_c$ vs. $x/L$, Eq.~(\ref{eta}).
In (b) and (c) ${L}=3{\pi\lambda_0}$ as in Fig.~\ref{fig3}. The parameter
${L}/{\pi\lambda}=\{100,10,3,1,0,-0.5,-0.7,-0.8\}$ from top to bottom.
}
 \label{fig4}
\end{figure}

In an open circuit, $j=0$, illustrated in Fig.~\ref{fig3}, linear terms
$\propto(1-x/L)$ in \eqref{phiPhi} and \eqref{eta}
cancel out [see Eq.~\eqref{CVC}]. Therefore,
\bea\nonumber 
&&\frac{\mu(x)-\mu(L)}{-e{\cal U}_c}={\cal
N}\left(\frac{\pi x}{L},\frac{L}{\pi\lambda}\right)
=\frac{\lambda_0}{{\cal U}_c \lambda}\phi(x,0)\\
&&=\frac{\eta(x)-\eta(L)}{(\lambda/\lambda_0-1)e{\cal
U}_c}\,,\qquad  j=0. \label{prop} 
\eea
General behavior (including $j\neq0$)
is illustrated in Fig.~\ref{fig4},
where we subtract the linear
term $\propto V(1-x/L)$ in
Eqs.~\eqref{phiPhi} and \eqref{eta}. The rest, expressed in
units  ${\cal U}_c$, is fully determined by two parameters
${L}/{\pi\lambda}$ and $\lambda/\lambda_0$. The relation between
$j$ and $V$ for an arbitrary measurement scheme follows from
CVC \eqref{CVC}.

%%%%%%%%%%%%%%%%%%%%%%%%%%%%%%%%%%%%%%%%%%%%%%%%%%%%%%%%%%%%%%%%%%%%%%%
%%%%%%%%%%%%%%%%%%%%%%%%%%%%%%%%%%%%%%%%%%%%%%%%%%%%%%%%%%%%%%%%%%%%%%%%%%%%

Several parametric regions deserve special attention.

(i) Continuous limit $\lambda\ll L$. This limit of strong
screening is illustrated by three upper curves (green and black) in
Figs.~\ref{fig4}a, b, and c. 
In the region $x,\lambda\ll L$, the
summation in Eq.~\eqref{N} can be replaced by integration,
\bea\nonumber 
&&{\cal N}\left(\frac{\pi
x}{L},\frac{L}{\pi\lambda}\right)
\simeq\frac{2}{\pi}\int\limits_{x/\lambda}^\infty\frac{d t}{t}\sin\left(t-\frac{x}{\lambda}\right)\\
&&= \frac{2}{\pi}{\rm ci}\frac{x}{\lambda}\sin\frac{x}{\lambda}
-\frac{2}{\pi}{\rm si}\frac{x}{\lambda}
\cos\frac{x}{\lambda},\qquad \lambda,x\ll L. \label{L>} 
\eea 
Here
${\rm si}$ and ${\rm ci}$ are the integral sine and cosine defined
by ${\rm ci}(z)+i\,{\rm si}(z)=-\int_z^\infty \left(e^{i
t}/t\right)\,dt$. Similar distribution was obtained in
Ref.~\onlinecite{Shik} for equilibrium semi-infinite 2DES 
with different boundary conditions at $B=0$. At large distances,
Eq.~(\ref{L>}) produces a power-law decay, 
\be\label{plaw} 
{\cal
N}\simeq 2\lambda/\pi x\,,\qquad \lambda\ll x\ll L, 
\ee 
characteristic for 2D electrostatics.

(ii) Enhanced screening, $\lambda\ll\lambda_0$, see two upper (green) 
curves in Figs.~\ref{fig4}a, b, and c. This case
corresponds to the well developed maxima of the microwave-induced
resistance oscillations, $\sigma\gg\sigma_0$. Beyond the screening
length, $x>\lambda$, both $\phi$ and $\eta$ change linearly with
$x$. The slope,
\be\label{slope} 
-\frac{\nabla_s \eta}{e}=-\nabla_s \phi(x,0)=\frac{V+{\cal
U}_c(1-\lambda/\lambda_0)}{L}=\frac{j}{2\sigma},
\ee 
is given by
Eqs.~\eqref{phiPhi} and \eqref{eta} with ${\cal N}\to 0$. The
second (diffusion) term in Eq.~\eqref{j} is relevant only in the
narrow region $x<\lambda$, where $|\nabla_s \mu/e|\simeq |\nabla_s
\phi|_{z=0}|\lambda_0/\lambda\gg|\nabla_s \phi|_{z=0}|$.

(iii) Equilibrium, $\lambda=\lambda_0$, see black curves (third from top)
 in Figs.~\ref{fig4}a, b, and c. When ${\cal U}_c\neq 0$
but there is no microwave field ($\lambda=\lambda_0$), 
$\phi(x,0)$ and $n_s(x)$ vary near the left contact on scale $\lambda_0$, while the
nonlinear part of $\eta(x)$ is zero. Since the
Einstein relation holds, two terms in Eq.~\eqref{j} combine into
one, $j=-2\sigma\nabla_s \eta/e=-2e\nu_0 D\nabla_s \eta=2\sigma_0
V/L$. In other words, two terms in Eq.~\eqref{j} partially
compensate each other such that $\nabla_s\eta=-eV/L$ remains
constant in space despite both $\nabla_s n_s$ and $\nabla_s e\phi$
vary strongly with $x\leq \lambda_0$.

(iv) Diffusion dominated screening, $\lambda>\lambda_0$, see orange curves 
(fourth from top) in Figs.~\ref{fig4}a, b, and c. In this regime, diffusion
dominates in the sense $e^2 \chi D>2\sigma$. 
The nonequilibrium screening charge distribution [which is smooth compared to the equilibrium case (iii)]
produces an ``overshoot'' in $\phi(x,0)$: two regions with the opposite
orientation of induced electric field appear, Fig.~\ref{fig4}b
(the total field may not change sign if the external voltage is
sufficiently large). The corresponding nonequilibrium correction
to $\eta$, Fig.~\ref{fig4}c, changes sign and remains negative at
$\lambda^{-1}<\lambda_0^{-1}$ (including negative $\lambda^{-1}$, see below).

(v) Zero-conductivity state, $\lambda^{-1}=0$, see blue curves 
(fourth from bottom) in Figs.~\ref{fig4}a, b, and c. In the limit
$L\to\infty$, homogeneous state of 2DES becomes electrically
unstable at $\lambda^{-1}=0$.\cite{Andreev} In the finite system, the instability
threshold shifts to negative $\sigma$ determined by the
condition $L/\pi\lambda=-1$. 

As was discussed in Sec.~\ref{enhancement}, in an open circuit, $j=0$,
the electrostatic potential $\phi(0,0)$ has a
singularity at $\sigma=0$, which implies the necessity to 
include the non-linear effects. This divergence does not appear
if one fixes the voltage $V$ instead of current. Indeed,
Eq.~\eqref{j} with $\sigma=0$ and boundary condition \eqref{BCn}
yields the charge density varying linearly with $x$, ${\cal N}=1-x/L$, 
see Fig.~\ref{fig4}a. The corresponding current density \eqref{j}
 has a $V$-independent value
\be\label{js}
j_s=e D(n_0-n_+)/L=-\chi e^2 D {\cal U}_c
\ee 
fixed by the boundary conditions.
The voltage $V$ decouples and can be arbitrary
within the range where the linear-response approximation
\eqref{j} is justified.

At $\lambda^{-1}=0$, second term in Eq.~(\ref{eta})
(which contains an indeterminate form of the type $\infty\cdot 0$)
can be represented as
\be
\label{eta0}
\frac{\eta(x)\!-\!\eta(L)\!-\!eV\left(1-\frac{x}{L}\right)}{e\,{\cal U}_c}=\frac{2L}{\pi^2\lambda_0}
{\rm Im}\, {\rm Li}_2(e^{%\frac
{i\pi x}/{L}}),
\ee
see Fig.~\ref{fig4}c. Here we used $\sum_{k=1}^\infty k^{-2}\sin \frac{\pi k x}{L}
={\rm Im}\; {\rm Li}_2(\exp[i\pi x/L])$,
where ${\rm Li}_2(z)$ is the dilogarithm function.
The potential profile induced by the linear variation of charge density ${\cal N}=1-x/L$ is
$e\phi(x,0)=\eta(x)-\eta(L) +e\,{\cal U}_c(1-x/L)$, Fig.~\ref{fig4}b.

(vi) Stable negative conductivity state, $-\pi/L<\lambda^{-1}<0$, 
see three lowest (red) curves in Figs.~\ref{fig4}a, b, and c.
The distributions \eqref{n}, \eqref{phiPhi}, and \eqref{eta}
are dominated by first harmonics in Eq.~\eqref{N},
\be\label{N1}
{\cal N}\left(\frac{\pi x}{L},\frac{L}{\pi\lambda}\right)\simeq
\frac{2}{\pi}\,\frac{\sin({\pi x}/{L})}{1+L/\pi\lambda},
\ee
which diverges at the instability threshold, $1+L/\pi\lambda=0$.

\subsubsection{Towards the domain structure}
\label{broken}

The divergence of our solution at the threshold $\lambda=-L/\pi$ signals
an instability and transition to the domain phase. We believe that inclusion of nonlinear
effects (in particular, taking into
account dependence of $\sigma$ entering Eq.~\eqref{j} on the electric field
$\nabla_s\phi$) should make
the theory applicable also in the domain regime $1+L/\pi\lambda<0$.
Work in this direction is currently underway.

Let us emphasize the emergence of a nonmonotonous profile of the
electrostatic potential [see Figs. 3 and 4(b)] implying formation
of two regions with opposite directions of the electric field. In
our solution, the direction of electric field ($+-$ vs. $-+$) in
these two regions is determined by the sign of ${\cal U}_{\rm c}$.
A non-zero ${\cal U}_c$ explicitly breaks the inversion symmetry
$L/2+x\leftrightarrow L/2-x$. We argue that this effect which
emerges at $\lambda<0$ and grows while the system approaches the
instability threshold, is the precursor of the domain structure
which fully develops at $\lambda^{-1}<-\pi/L$. As was discussed in
Sec.~\ref{homo}, in the homogeneous case ${\cal U}_c=0$ the
inversion symmetry is preserved at $\sigma_c<\sigma<0$ and gets
spontaneously broken at the instability threshold
$\sigma_c=\sigma$.

A further important direction for future work is to go beyond the
mean field approximation and to study the effects of fluctuations
and noise on the transition into the ZRS regime. It is expected
that field--theoretical approaches developed for nonequilibrium
phase transitions, in particular, Martin-Siggia-Rose formalism,
will be useful in this respect. First steps in this direction were
made in Ref.~\onlinecite{Balents05}.

\section{Summary}

In summary, the effect of the microwave
radiation on the electron energy distribution function of a 2DES causes
giant magneto-oscillations of the conductivity relative to its
dark value and practically does not alter the electron diffusion
coefficient. Such effect leads to magneto-oscillations of the
photo-galvanic signals and of the screening length, which affects
the potential profile in a sample. At
the oscillation minima, the conductivity can become negative,
which leads to a negative value of the non-equilibrium screening length. We have
derived the stability condition at which a finite 2DES can possess
a stable state with a negative conductivity.
When the
conductivity becomes negative, the differential resistance
and the photo-voltage also change their signs, which can be
observed experimentally.
We have further solved the combined transport and electrostatic problem
and determined the profiles of the potential and the electron density
inside the sample.
In the stable state with a negative conductivity,
the potential profile consists of two regions with opposite
directions of the electric field.
The amplitude of these fields increases when the system approaches the
instability threshold. This effect is a precursor of the domain structure
in the regime of spontaneously broken symmetry.

This work was supported by the Deutsche Forschungsgemeinschaft and by
the Russian Foundation for Basic Research.  S.I.D. gratefully acknowledges
 fruitful discussions with Yu.~A.~Bychkov.

\end{document}